\documentclass[fleqn,usenatbib]{mnras}

\usepackage{newtxtext,newtxmath}
\usepackage{balance}

\usepackage[T1]{fontenc}

\DeclareRobustCommand{\VAN}[3]{#2}
\let\VANthebibliography\thebibliography
\def\thebibliography{\DeclareRobustCommand{\VAN}[3]{##3}\VANthebibliography}

\usepackage{graphicx}	
\usepackage{amsmath}	
\newcommand{\angstrom}{\text{\normalfont\AA}}
\usepackage{subfigure} 
\graphicspath{{./}{figures/}}

\title[Seyfert NGC\,2639]{AGN Feedback Through Multiple Jet Cycles in the Seyfert Galaxy NGC\,2639}

\author[V. Rao et al.]{Vaishnav V. Rao,$^{1}$%
\thanks{E-mail: \href{mailto:vvrao@umich.edu}{vvrao@umich.edu}}%
P. Kharb,$^{2}$
Rubinur K.,$^{3}$
Silpa S.,$^{2}$
N. Roy,$^{4}$
B. Sebastian,$^{5}$
V. Singh,$^{6}$
J. Baghel,$^{2}$
S. Manna,$^{2}$ 
\newauthor{C. H. Ishwara-Chandra$^{2}$}
\\ \\
$^{1}$Indian Institute of Technology Bombay, Powai, Mumbai 400076, India\\
$^{2}$National Centre for Radio Astrophysics (NCRA) - Tata Institute of Fundamental Research (TIFR), S. P. Pune University Campus, Ganeshkhind, Pune 411007, India\\
$^{3}$Institute of Theoretical Astrophysics, University of Oslo, P.O box 1029 Blindern, 0315 OSLO, Norway\\
$^{4}$Johns Hopkins University, Department of Physics \& Astronomy, 3400 N. Charles Street, Baltimore, MD 21218, USA\\
$^{5}$Department of Physics and Astronomy, University of Manitoba, Winnipeg, MB R3T 2N2, Canada\\
$^{6}$Astronomy and Astrophysics Division, Physical Research Laboratory, Ahmedabad 380009, India
}

\pubyear{2022}
\begin{document}
\label{firstpage}
\pagerange{\pageref{firstpage}--\pageref{lastpage}}
\maketitle

\begin{abstract}
The Seyfert galaxy NGC\,2639 was known to exhibit three episodes of AGN jet/lobe activity. We present here the upgraded Giant Metrewave Radio Telescope (uGMRT) 735~MHz image of NGC\,2639 showing a fourth episode as witnessed by the discovery of $\sim9$~kpc radio lobes misaligned with the previously known $\sim1.5$~kpc, $\sim360$~parsec, and $\sim3$~parsec jet features detected through the Karl G. Jansky Very Large Array (VLA) and the Very Long Baseline Array (VLBA), respectively. Using the spectral ageing software BRATS, we derive the ages of the $\sim9$~kpc, $\sim1.5$~kpc, and $\sim360$~parsec episodes to be, respectively, $34^{+4}_{-6}$ Myr, $11.8^{+1.7}_{-1.4}$ Myr, and $2.8^{+0.7}_{-0.5}$ Myr, and conclude that minor mergers occurred $9-22$ Myr apart. NGC\,2639 shows a deficit of molecular gas in its central $\sim6$~kpc region. The GALEX NUV image also shows a deficiency of recent star-formation in the same region, while the star formation rate (SFR) surface density in NGC\,2639 is lower by a factor of $5-18$ compared to the global Schmidt law of star-forming galaxies. This makes NGC\,2639 a rare case of a Seyfert galaxy showing episodic jet activity and possible signatures of jet-driven AGN feedback.
\end{abstract}


\section{Introduction} \label{sec:intro}
Active galactic nuclei (AGN) are the high-luminosity, energetic centres of galaxies that are dominated by light emitted from matter accreting onto a supermassive black hole \citep[SMBH, M$_\mathrm{BH}\sim 10^6-10^9~\mathrm{M}_{\sun}$;][]{Rees1984,Padovani2017}. In a small fraction of AGN ($\sim10\%$), relativistic jets are launched from the SMBH up to hundreds of kpc to even Mpc scales \citep{Readhead1978,Orr1982,Heckman2014}. AGNs have been historically broadly classified into Seyfert galaxies and quasars, with the fundamental difference lying in their bolometric luminosity (L$_\mathrm{bol}$) with L$_\mathrm{bol} \leq 10^{12}$~L$_{\sun}$ for Seyfert galaxies and higher for quasars \citep{Schmidt1983,Soifer1987}. Based on the presence or absence of broad bases to permitted emission lines in the optical spectra, Seyfert galaxies are further classified into types 1 and 2 \citep{Khachikian1974,Hickox2018}. It is widely believed that different viewing angles of the central engine surrounded by an obscuring torus explain the two Seyfert classes, with type 1s being viewed nearly face-on, although there are several exceptions \citep{Antonucci1993,Ho2008,Netzer2015}. Seyfert galaxies are primarily classified as `radio-quiet' (RQ) AGN. Based on 6~GHz VLA observations of SDSS quasi-stellar objects, \citet{Kellermann2016} have suggested RQAGN to be those having $21 \le \mathrm{log[L_6 (W~Hz^{-1})] \le 23}$. 

Outflows in Seyfert galaxies are not a well-understood phenomenon \citep{Panessa2019}. Only sensitive radio observations reveal the presence of radio structures with typical extents $\sim10$~kpc in Seyfert galaxies. \citet{Gallimore2006, Singh2015} found that $\gtrsim40\%$ of Seyfert galaxies in statistically large samples exhibit kiloparsec-scale radio structures (KSRs). It is unclear how these radio outflows are generated. The contribution of the AGN versus star formation to the radio emission is being debated \citep[e.g.,][]{Baum1993, Hota2006,Sebastian2020}. Several authors \citep{Malzac2001, Markoff2005, Kharb2015, Behar2015, Wong2016} have argued the case for Seyfert outflows being outflowing accretion-disk coronae. High-resolution VLBI observations, on the other hand, have detected the presence of parsec-scale jets in a majority of radio-bright Seyfert galaxies \citep[e.g.,][]{Thean2000, Nagar2005, Kharb2010, Mezcua2014, Baldi2018, Kharb2021}, making a strong case for Seyfert galaxies being jetted AGN. \cite{Ho2008} have argued that small collimated outflows in low luminosity AGN like Seyferts can be agents of radiative or mechanical AGN feedback and this energy injection can lead to unsteady, intermittent accretion with a short duty cycle.

\begin{figure*}
\centerline{
\includegraphics[width=12cm]{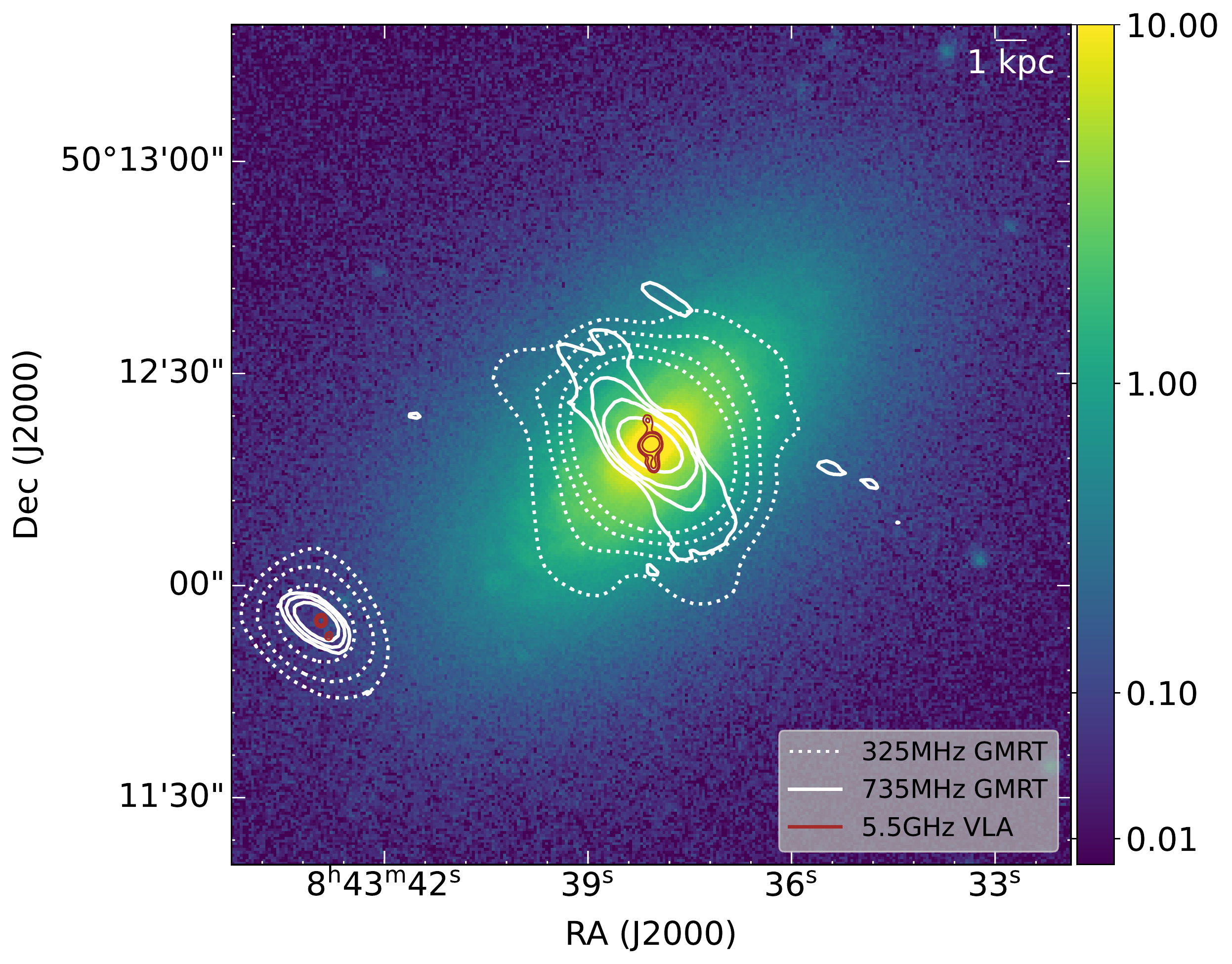}}
\caption{
A radio-optical overlay of NGC\,2639. The optical image comes from SDSS r-band. The dotted, white contours represent the 325~MHz GMRT data with contour levels $(1,2,4,8)\times1.6$~mJy~beam$^{-1}$. The solid white contours represent the 735~MHz GMRT data with contour levels $(1,2,4,32)\times0.6$~mJy~beam$^{-1}$. The solid brown contours come from the VLA 5.5~GHz image with contour levels $(2,4,8,64)\times 0.03$~mJy~beam$^{-1}$.}
\label{fig:radopt}
\end{figure*}

AGN are believed to regulate galaxy growth by injecting energy into the surrounding gas which has the effect of either heating and/or expelling star-forming material (``negative feedback'') or facilitating localized star-formation (``positive feedback'') \citep{AlexanderHickox2012,Fabian2012,Morganti2017,Harrison2017}. While AGN ``feedback'' is believed to be a fundamental process of galaxy formation, there are many outstanding questions from an observational point of view. 
AGN feedback has been suggested to come in the ``quasar mode'' or the ``maintenance/jet mode'' \citep[e.g.,][]{Croton2009,Bower2012}. The former is associated with radiatively driven AGN, where the energetic photons coupled with the surrounding gas drive high-velocity winds through the host galaxy and result in the removal of the star-forming material \citep[e.g., ][]{Faucher-GiguereQuataert2012, Costa2018}, while the latter is associated with AGN hosting radio jets that can transfer mechanical power (jets could heat, shock or entrain gas) and regulate star formation \citep[e.g.,][]{McNamaraNulsen2012,Mahony2013,Morganti2017,HardcastleCroston2020}. While highly collimated jets might not be efficient agents of AGN ``feedback'', presumably due to the smaller working surfaces at their advancing ends, relatively isotropic impacts via changes in jet direction can be highly effective \citep{King2015}. Moreover, the contribution of jets towards the total AGN energy budget has been suggested to be $\lesssim20$\% \citep{Cattaneo2009,Merloni2008,Shankar2008}.

\citet{Sanders1984} have argued that individual Seyfert activity episodes typically have a shorter duration than the minimum statistical lifetime of Seyfert activity in a particular galaxy ($3-7\times 10^8$ yr), which would imply that the nuclei evolve through at least a 100 recurring Seyfert episodes. If it is assumed that Seyfert episodes in a particular galaxy are due to accretion onto the central black hole, the short lifetime of Seyfert events would require separate episodes to be caused by distinct accretion events. Recent studies from the SDSS, though, have suggested that accretion rate changes are common within a given Seyfert duty cycle, producing a much more complex picture of accretion in Seyfert galaxies \citep{Koulouridis2014,Elitzur2014,Koulouridis2016}. So far, only a handful of Seyfert galaxies such as Mrk~6 \citep{Kharb2006} and NGC~2992 \citep{Irwin2016} have been known to exhibit multiple kpc-scale radio lobes, with Mrk~6 showing three jet episodes. Here and in other cases discussed in this paper, different jet ``episodes'' are defined by surface brightness discontinuities, with each jet episode being delineated by the presence of either terminal hotspot-like features or lobes with well-defined edges i.e., surface brightness discontinuity of the edges is at least three times the \textit{rms} noise. \citet{Saikia2022} have pointed out that different jets are also characterized by significant differences in position angle (PA; measured here from north through east with north being at $0\degr$) and differences in the steepness of the spectral index.

Interestingly, \cite{Sebastian2020} found tentative evidence for multiple episodes in a majority (5/9) of their Seyfert galaxy sample. This fraction was higher than that observed in radio galaxies \citep[$\sim10-15\%$;][]{Jurlin2020} and was consistent with the theoretical expectation of \cite{Sanders1984}. Therefore, the rarity of known Seyfert galaxies with episodic activity may not be due to a true absence but rather due to the difficulty in their identification due to the low surface brightness of their lobes, small spatial extents, lack of collimation, and confusion with the radio emission from star formation. Hence, it is essential to design methods to identify various episodes of jetted emission in Seyfert galaxies to truly understand the life cycle of jets in these systems and their impact on the environment. Multi-frequency, multi-scale/resolution observations are one such method. Although several AGN, including radio galaxies and Seyfert galaxies, have been shown to exhibit two episodes of jet activity \citep[see][and references therein]{Saikia2009, Shabala2020, Saikia2022}, it is scarce to find sources with three or more episodes \citep[although see the simulations of][]{Lalakos2022}. Here we report the discovery of a Seyfert galaxy NGC\,2639, which shows four AGN jet activity episodes. 

NGC\,2639 (a.k.a. UGC\,4544) is a Seyfert 2 galaxy \citep[][see Figure~\ref{fig:radopt}]{Lacerda2020}. At its redshift of 0.01113 (luminosity distance 48.2 Mpc), $1\arcsec$ corresponds to 0.229~kpc. NGC\,2639 hosts water vapour megamasers most likely in a cool dense nuclear disk \citep{Wilson1995}. The maser in this galaxy has a luminosity of $71$ L$_{\odot}$ and has a prominent emission component of 3300 km s$^{-1}$ (width 5.4 km s$^{-1}$) surrounded by some weaker emission components \citep{Braatz1994}. \citet{Berrier2013} have estimated the mass of the black hole (BH) in NGC\,2639 using the M$_\mathrm{BH}-\sigma$ relation \citep{Ferrarese2000}, which turns out to be M$_\mathrm{BH} = (1.48 \pm 0.43)\times10^8$~M$_{\odot}$. A more accurate BH mass is not yet available from the water vapor megamaser data due to the non-detection of ``satellite'' lines, as described by \citet{Wilson1995}. The stellar mass of the host galaxy has been estimated to be $1.48\times10^{11}$~M$_\odot$ by \citet{Sweet2018}. The Eddington ratio for the BH in NGC\,2639 is obtained to be $\frac{L_\mathrm{bol}}{L_\mathrm{Edd}}= 4\times 10^{-4}$, where L$_\mathrm{bol}$ is the bolometric luminosity and L$_\mathrm{Edd}$ is the Eddington luminosity. L$_\mathrm{bol}$ was obtained using the $2-10$~keV absorption-corrected X-ray luminosity L$_\mathrm{X}$ $=46.6\times10^{40}$~erg~s$^{-1}$ for NGC\,2639 \citep{Terashima2002}, and was estimated via the relation $\mathrm{L_{bol} = 15.8~L_X}$ \citep[see][]{Ho2009}. The Eddington luminosity was obtained through L$_\mathrm{Edd}=1.26\times10^{38}\left(\frac{\mathrm{M_{BH}}}{\mathrm{M_\odot}}\right) = 1.86\times10^{46}$~erg~s$^{-1}$. The extreme dimness of the nucleus, radiating well below the Eddington limit strongly suggests that the AGN accretion flow is radiatively inefficient in NGC\,2639 \citep{Heckman2014}.

Using multi-resolution radio observations with the Karl G. Jansky Very Large Array (VLA) and the Very Long Baseline Array (VLBA), \citet{Sebastian2019} reported three pairs of radio jets/lobes in NGC\,2639. The three lobes are misaligned with each other without any apparent signatures of continuity. \citet{Sebastian2019} concluded that these three lobes were representative of three distinct episodes of AGN jet activity and ruled out other scenarios, including multiple jets from independent AGN, slow jet precession, and jet bending due to pressure gradients within the galaxy. The presence of a jet precession was discounted based on the lack of supporting evidence for the continuity in flux densities or trajectories across the three distinct lobe pairs. Similarly, the likelihood of multiple independent jets originating from different AGNs was deemed implausible due to their extremely low probabilities. Lastly, pressure gradients were eliminated as a possibility due to the observed misalignment of the three lobes from the minor axis, which is incompatible with this hypothesis. However, it is worth noting that \citet{Xanthopoulos2010} have favored jet-medium interaction rather than jet precession to explain the S-shaped, east-west MERLIN jet in NGC\,2639 (corresponding to the 360~parsec VLA jet; see ahead).

\begin{figure*}
\centering{
\includegraphics[width=18cm]{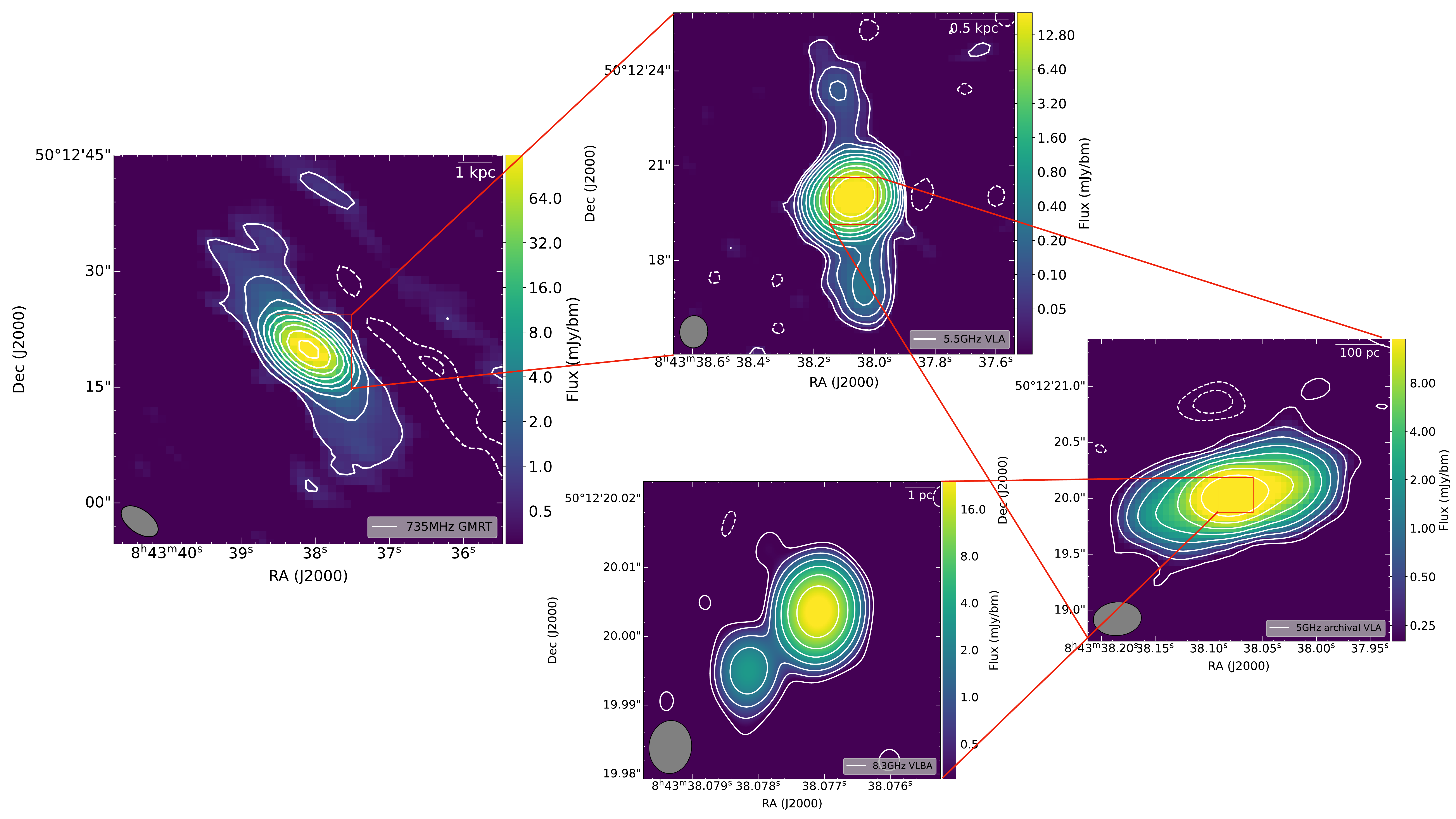}}
\caption{The four AGN jet episodes of NGC\,2639. (Left) 735~MHz GMRT total intensity image. The $\sim9$~kpc northeast-southwest radio lobes at PA=$34\degr$ are seen in this image. Contour levels: $(-2,-1,1,2,4,8,16,32,64,128,256)\times0.6$~mJy~beam$^{-1}$. Beam at the bottom left corner is of size: $5.48\arcsec\times3.0\arcsec$ at PA=$54.6\degr$. (Top) 5.5~GHz VLA total intensity image. Contour levels: $(-2, -1, 1, 2, 4, 8, 16, 32, 64, 128, 256, 512)\times 0.03$~mJy~beam$^{-1}$. The $\sim1.5$~kpc north-south radio jets at PA=$6\degr$ are seen here. Beam size: $1.02\arcsec\times0.89\arcsec$ at PA=$-5.8\degr$. (Right) 5~GHz VLA radio image. Contour levels: $(-2,-1, 1, 2, 4, 8, 16, 32, 64, 128)\times 0.164$~mJy~beam$^{-1}$. The $\sim360$ parsec east-west lobes at PA=$106\degr$ are seen in this image. Beam size: $0.43\arcsec\times0.30\arcsec$ at PA=$-85.4\degr$. (Bottom) 8.3 GHz VLBA image showing a $\sim3$~parsec jet at PA = $130\degr$. Contour levels: $(-2,-1, 1, 2, 4, 8, 16, 32, 64)\times0.239$~mJy~beam$^{-1}$. Beam size: $7.7$ mas$\times6.2$ mas at PA=$-4.9\degr$.}
\label{fig:insets}
\end{figure*}

The jet extents of the three episodes were 1.5~kpc, 360~parsec, and 3~parsec as revealed by the VLA image at 5.5~GHz, historical VLA image at 5~GHz, and VLBA image at 8.3~GHz, respectively (see Figure \ref{fig:insets} and Table~\ref{tab:Table1}). Following the criterion of \citet{Kellermann2016}, NGC\,2639 is a radio-quiet AGN having $\mathrm{log[L_6 (W~Hz^{-1})] = 22.15}$, obtained from its 5.5~GHz VLA lobes using a spectral index of $-0.6$. In this paper, we present new upgraded Giant Metrewave Radio Telescope (uGMRT; hereafter GMRT) images of NGC\,2639 which reveal an additional set of radio lobes, not previously detected at GHz frequencies (see Figure~\ref{fig:radopt}). The paper is divided as follows. In Section~\ref{sec:radio}, we present the radio data analysis, followed by the spectral ageing analysis in Section~\ref{sec:brats}. The results and discussion follow in Sections~\ref{sec:results} and \ref{sec:discussion}, respectively, while the conclusions follow in Section~\ref{sec:summary}. In this paper, we have adopted the following cosmological parameters: $H_0 = 73$ km s$^{-1}$ Mpc$^{-1}$, $\Omega_\mathrm{mat}= 0.27$, $\Omega_\mathrm{vac}=0.73$. Throughout the paper, spectral index $\alpha$ is defined such that flux density $S_\nu \propto \nu^\alpha$.

\section{Radio Observations and Data Analysis}\label{sec:radio}
NGC\,2639 was observed with the GMRT (Project code: 39\textunderscore090) at 735~MHz (band 4) on November 15, 2020. The observing band consisted of a single spectral window, ranging from 550~MHz to 950~MHz, across 4096 channels. The total on-source time was $\approx$36 minutes. 3C\,147 was used as the amplitude calibrator and 0834+555 as the phase calibrator. Initial data editing and calibration were carried out using the \texttt{CAPTURE} continuum imaging pipeline for GMRT \citep[][]{capture2021} on \texttt{CASA} \citep[version 6.4;][]{CASA}. The multi-term multi-frequency synthesis \citep[\texttt{MT-MFS}; see ][for more details]{Rau2011} algorithm with two Taylor terms was used while imaging in \texttt{CASA} to account for wide-band related errors while deconvolving. Four rounds of phase-only self-calibration followed by four rounds of A\&P (amplitude and phase) self-calibration were performed before the final image of NGC\,2639 was created using the \texttt{tclean} task in \texttt{CASA}. \texttt{tclean} is the radio interferometric image reconstruction task that contains standard ``clean'' based algorithms \citep[e.g.,][]{Hogbom1974,Clark1980} along with algorithms for multi-scale and wideband image reconstruction. The final \textit{rms} noise in the resulting image was $\sim90~\mu$Jy beam$^{-1}$.

Calibrated data of NGC\,2639 used by \citet{Sebastian2019} from the VLA B-array at 5.5 GHz, were was also available. The data was imaged using the \texttt{tclean} task in \texttt{CASA} using similar steps as above. Images obtained from archival VLA data at 1.5~GHz from 1985 and GMRT data at 325~MHz were used along with the above data sets for a spectral-ageing analysis. In addition to this, VLA images at 5~GHz from 1998 and VLBA images at 8.3~GHz from 2011, which were analyzed by \citet{Sebastian2019}, were included in our analysis as well. These figures are included as insets in Figure~\ref{fig:insets}. Tables~\ref{tab:Table1} and \ref{tab:Table2} include a summary of the datasets used along with their corresponding source properties. The flux densities were estimated using the \texttt{CASA} Viewer. We made rectangular boxes around the extended emission using the drawing tool and noted the flux density values from the statistics tab of the region manager for the selected regions. The flux density errors were obtained through the formula $\sqrt{(\sigma_\mathrm{rms}\cdot\sqrt{N_b})^2 + (\sigma_p\cdot S)^2}$, where $\sigma_\mathrm{rms}$ is the \textit{rms} noise in the image, $N_b$ is the number of beams that span the source area, $\sigma_\mathrm{p}$ is the percentage error on the absolute flux density scale (assumed to be a conservative 0.1 for the GMRT, VLA{\footnote{\url{https://science.nrao.edu/facilities/vla/docs/manuals/oss/performance/fdscale}}} and VLBA) and $S$ is the source flux density \citep{Kale2019}.

\begin{table*}
\caption{Details of the different datasets used for analysis in this paper} \label{tab:Table1}
\begin{tabular}{cccccl}
\hline
{Telescope} & {Project Code} & {Observing date} & {Frequency} & {Array} & {Beam size} \\
\hline
$^*$GMRT & 39\_090 & 2020-11-15 & 735~MHz & - & $5.48\arcsec \times 3.0\arcsec$, PA $54.6\degr$ \\
$^*$GMRT & 22\_002 & 2012-06-09 & 325~MHz & - & $12.1\arcsec \times 9.8\arcsec$, PA 37.6$\degr$ \\
$^*$VLA & 17B-074 & 2017-11-04 & 5.5~GHz & B & $1.02\arcsec \times 0.89\arcsec$, PA $-5.8\degr$ \\
VLA & GL022 & 1998-02-17 & 5.0~GHz & A & $0.43\arcsec \times 0.30\arcsec$, PA $-85.4\degr$ \\
$^*$VLA & AW126 & 1985-02-04 & 1.5~GHz & A & $1.45\arcsec \times 1.12\arcsec$, PA 51.9$\degr$ \\
VLBA & BC196J & 2011-03-12 & 8.3~GHz & - & 7.7 mas $\times$ 6.2 mas, PA $-4.9\degr$ \\
\hline
\end{tabular}

$^*$ indicates the datasets that have been used for the BRATS spectral ageing analysis.
\end{table*}

\begin{table*}
\caption{Source parameters in the different datasets used in the paper} \label{tab:Table2}
\begin{tabular}{cccccc}
\hline
{Telescope} & {Frequency} & {Peak Source Intensity (mJy~beam$^{-1}$)} & {\textit{rms} noise ($\mu$Jy~beam$^{-1}$)} & {Total Source Intensity (mJy)} & {Source PA}\\
\hline
GMRT & 735~MHz & 198 & 90 & $223 \pm 22$ & 34\degr\\
GMRT & 325~MHz & 249 & 128 & $286\pm29$ & 40\degr\\
VLA & 5.5~GHz & 42 & 9 & $57\pm6$ & 6\degr\\
VLA & 5.0~GHz & 54 & 54 & $79\pm8$ & 106\degr\\
VLA & 1.5~GHz & 61 & 82 & $93\pm 9$ & 180\degr\\
VLBA & 8.3~GHz & 29 & 75 & $33\pm3$ & 130\degr\\
\hline
\end{tabular}
\end{table*}

\section{Spectral Ageing Analysis using BRATS}\label{sec:brats}
We used the Broadband Radio Analysis ToolS (BRATS) Software \citep[][]{Harwood2013,Harwood2015} to carry out a spectral ageing analysis of the different jet episodes in NGC\,2639. We estimated the `minimum energy' magnetic field ($B_\mathrm{min}$) in the radio lobes assuming the equipartition of magnetic field and particle energy densities \citep[e.g.,][]{Pacholczyk1970} and used this as the magnetic field input in BRATS. We used the relations provided in \citet{O'Dea-Owen1986} to calculate $B_\mathrm{min}$. The volume filling factor, $\phi$, and the ratio of energy density of ions to electrons, $\eta$, was assumed to be unity. The upper and lower radio-band frequency cutoffs, $\nu_u$ and $\nu_l$ were taken to be 10~MHz and 100~GHz, respectively. The lobes were assumed to be cuboidal, with the volume estimated using the relation $l\times w\times w$, where $l$ and $w$ are the lengths and widths of the full double-sided lobes respectively. The input parameters, $B_\mathrm{min}$, and other estimates are noted in Table~\ref{tab:Table3}.

\subsection{GMRT North-East South-West Lobes}\label{sec:gmrt-spec-age}
To perform a spectral age analysis using BRATS, a minimum of three images of the same size, resolution, and beam size at different frequencies are needed. From the calibrated GMRT band-4 data of NGC\,2639, two subband images at central frequencies 643 MHz and 810 MHz were obtained using \texttt{tclean} with similar parameters as section~\ref{sec:radio}. These two images were smoothed and regridded using \texttt{imsmooth} and \texttt{imregrid} in \texttt{CASA} to match the resolution, image size, and beam size of the lower resolution 325 MHz archival GMRT image. The final common restoring beam used was 12.1$\arcsec\times9.8\arcsec$ at PA = 37.6$\degr$. In doing so, it was assumed that the lower resolution 325 MHz GMRT image contained the unresolved lobes which were seen at 735 MHz. The Jaffe-Perola (JP) model was used for fitting the synchrotron spectrum of the three radio images on BRATS \citep[see ][for details]{Harwood2013,Harwood2015}. A $5\sigma$ detection threshold was used with only the source regions selected. The injection index was chosen as $-0.2$ as this was the approximate spectral index of the core. In doing so, it was assumed that electrons are being accelerated near the center as opposed to the progressing edges of the lobes as in FRII radio galaxies \citep[e.g.,][]{Mahatma2020}. This was because no hotspots were observed in any of the radio images and the source resembles an FRI-type or Seyfert-like jet. The magnetic field (\texttt{bfield}) was set to the equipartition value of $7.6\times 10^{-6}$ G and the spectral age map was obtained using the \texttt{fitjpmodel} task on BRATS with \texttt{levels=5} and \texttt{Myears=60}.

\subsection{VLA North-South Lobes}\label{sec:vla-spec-age}
From the calibrated 5.5 GHz full-band VLA data of NGC 2639, two subband images at central frequencies 5 GHz and 6 GHz were obtained using \texttt{tclean}. These two images were smoothed and regridded using \texttt{imsmooth} and \texttt{imregrid} on \texttt{CASA} to match the resolution, image size, and beam size of the lower resolution 1.5 GHz archival VLA image, in which only the southern lobe was visible. The final common restoring beam used was $1.45\arcsec\times1.12\arcsec$ at PA $=51.9\degr$. With the core and southern lobe regions selected and a $5\sigma$ detection threshold, the synchrotron spectrum of the radio images were fitted with the JP model on BRATS. The injection index was again chosen to be $-0.2$ and the magnetic field was taken as $1.2\times 10^{-5}$ G from the equipartition estimate. The spectral age map was obtained from the \texttt{fitjpmodel} task with \texttt{levels=5} and \texttt{Myears=20}.

\begin{figure*}
\centering
    \subfigure[Spectral Age Map]{\includegraphics[height=5cm]{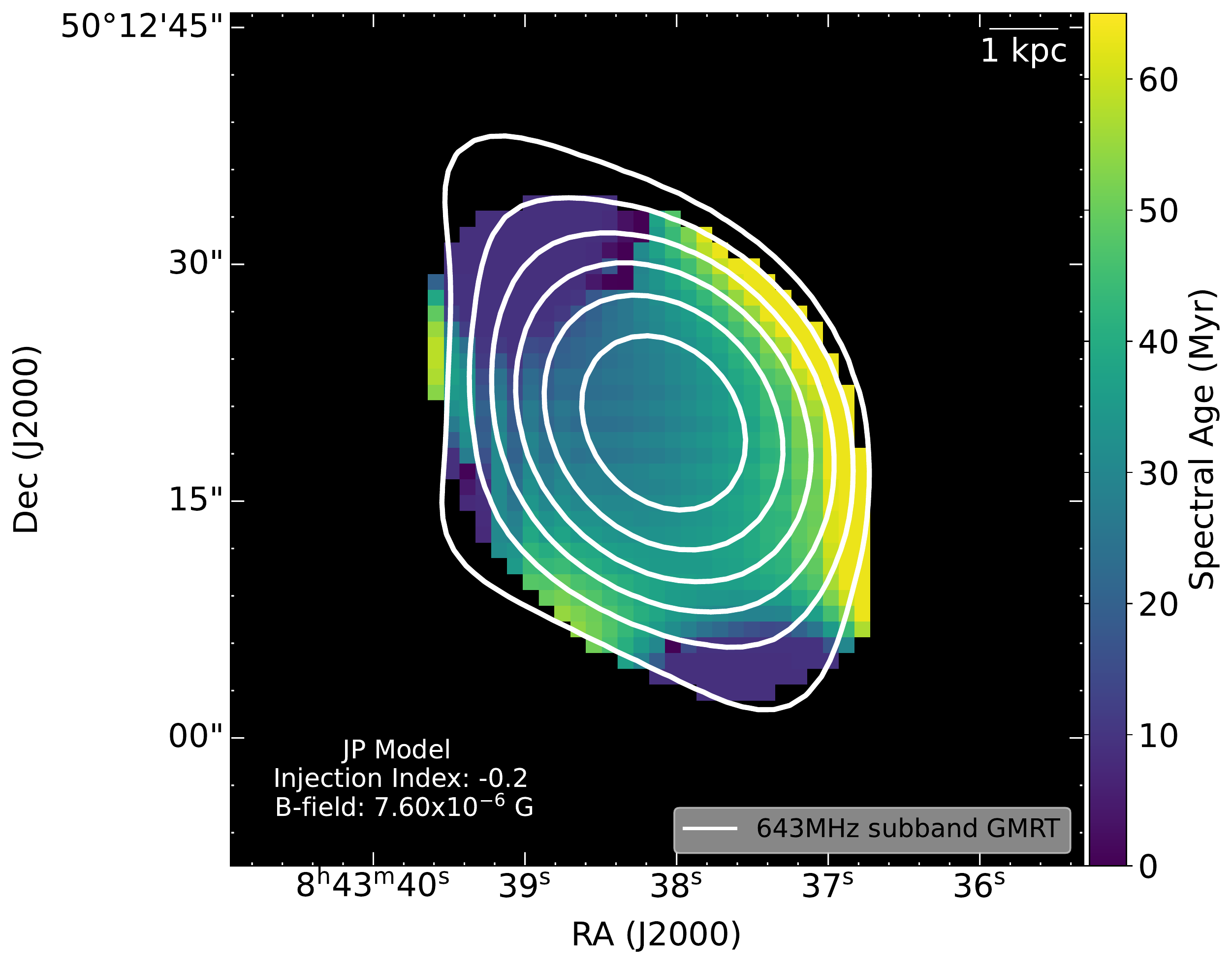}}
    \hfill
    \subfigure[$\chi^2$ map]{\includegraphics[height=5cm]{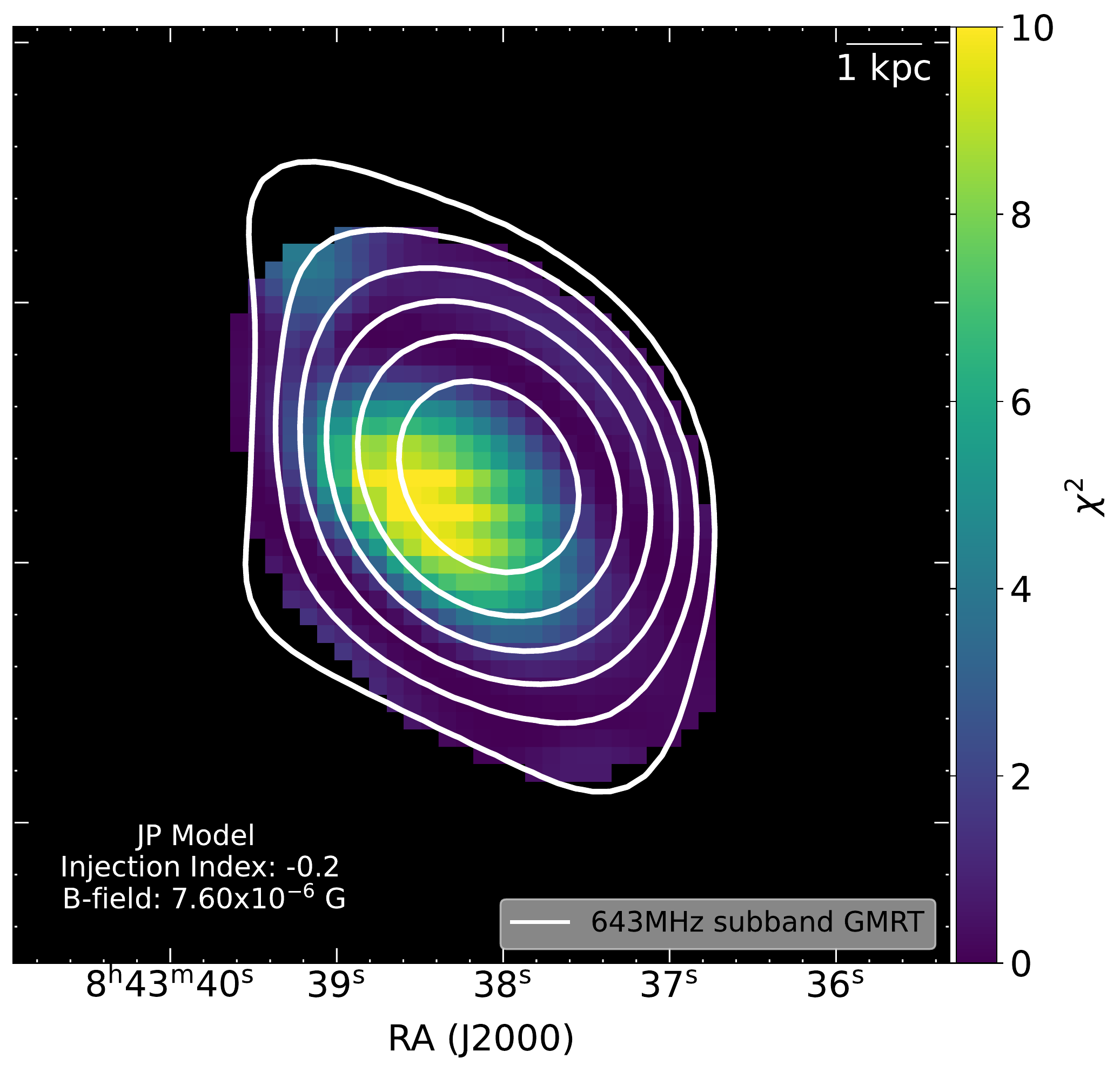}}
    \hfill
    \subfigure[Error map]{\includegraphics[height=5cm]{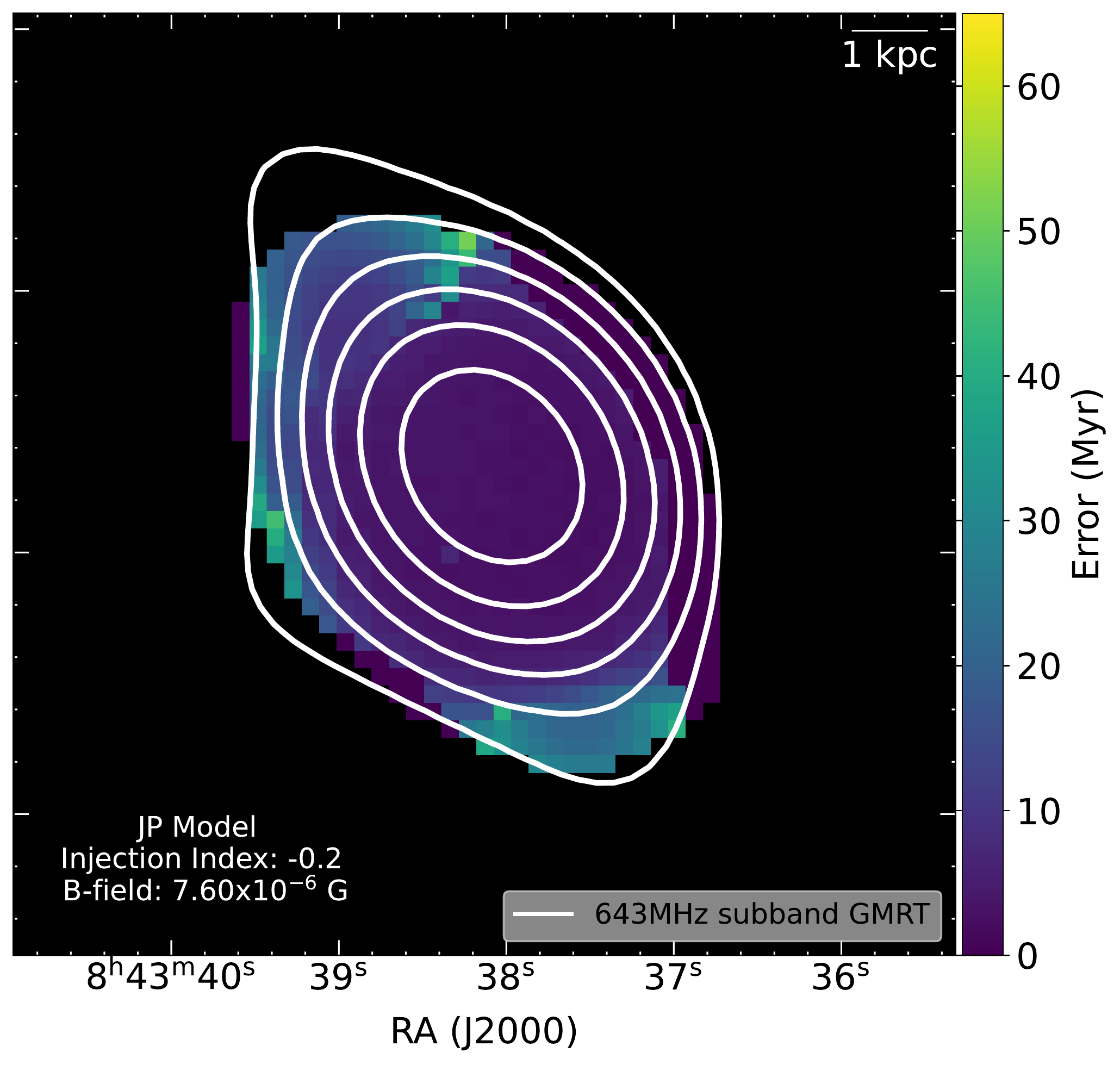}}
\caption{(Left) Spectral age map of the GMRT $\sim9$ kpc lobes obtained using the smoothed and regridded images at 325, 643, and 810 MHz. (Centre) The $\chi^2$ map of the fit to the JP model. (Right) The map of the error in estimated spectral age. The mean spectral age of the jet is $34^{+4}_{-6}$ Myr (Note: The purple patches at the top and bottom of the spectral age map along, the central AGN region, and the yellow sliver on the right of the map were excluded from calculating mean due to poor fitting and high errors as can be seen in the central and right panels). Contour levels are of the 643~MHz GMRT subband image (see section \ref{sec:gmrt-spec-age}) at: $(1,2,4,8,16,32,64)\times3.2$~mJy~beam$^{-1}$.}
\label{fig:gmrt-age}
\end{figure*}

\begin{figure*}
\centering
    \subfigure[Spectral Age Map]{\includegraphics[height=5cm]{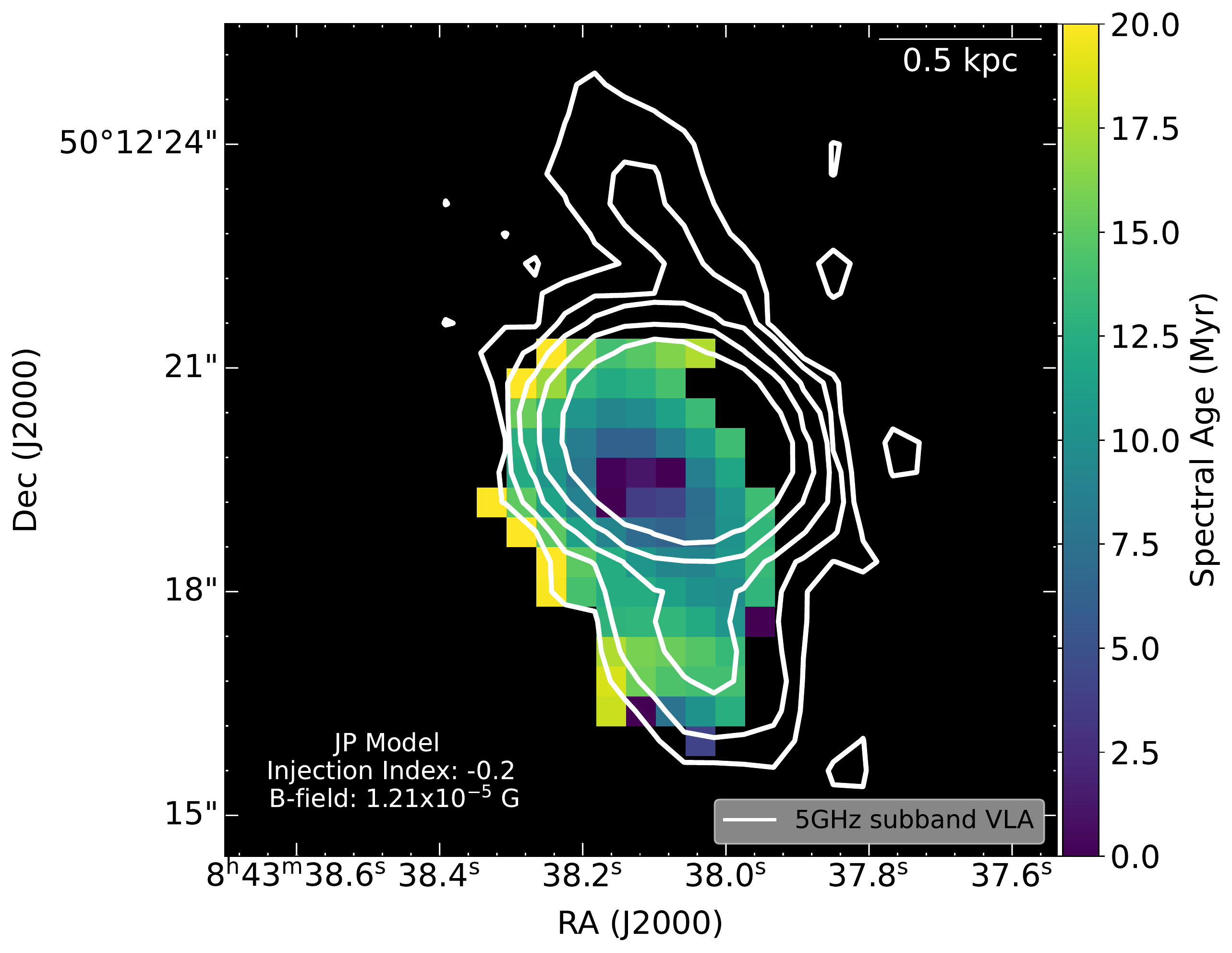}}
    \hfill
    \subfigure[$\chi^2$ map]{\includegraphics[height=5cm]{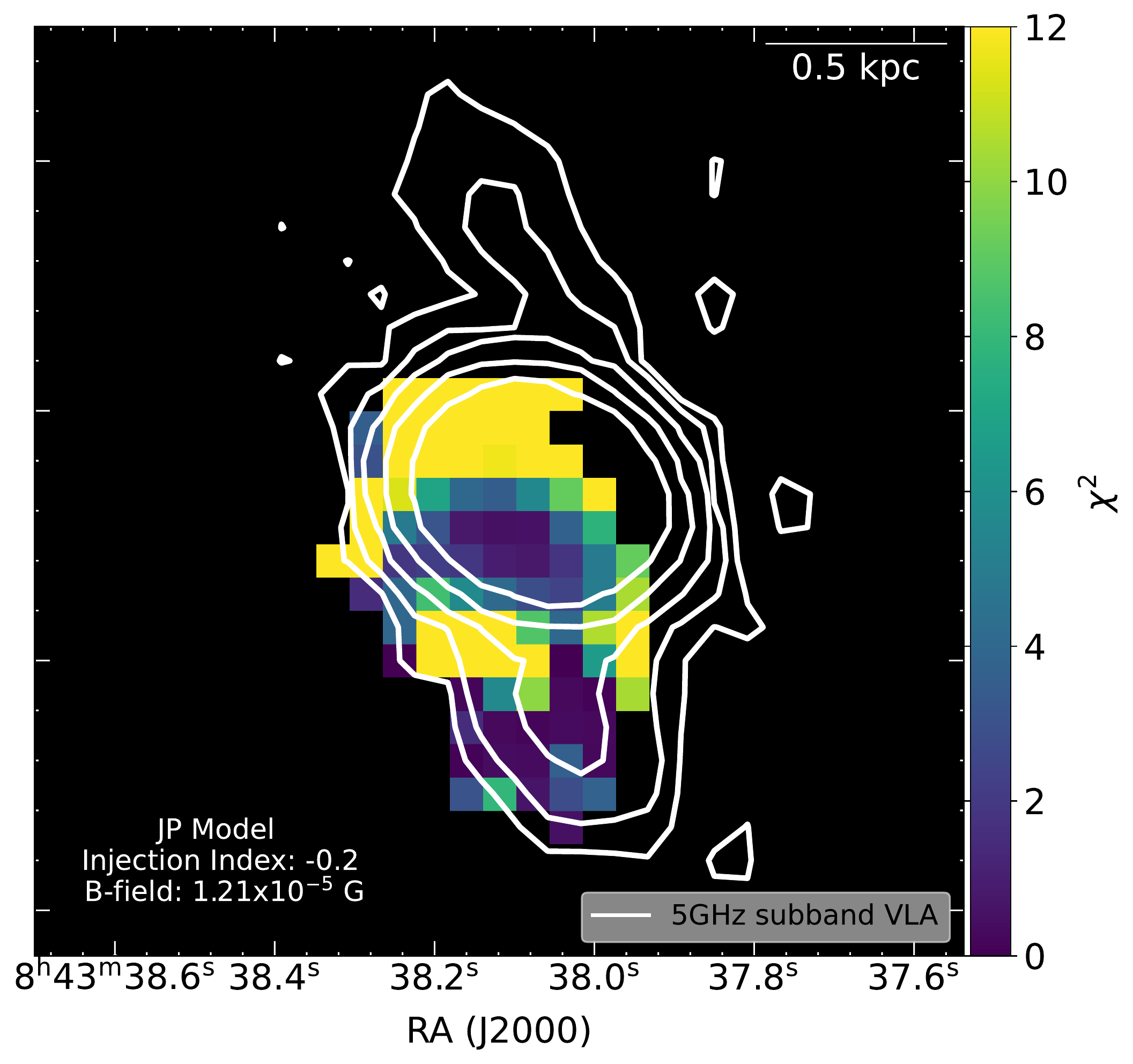}}
    \hfill
    \subfigure[Error map]{\includegraphics[height=5cm]{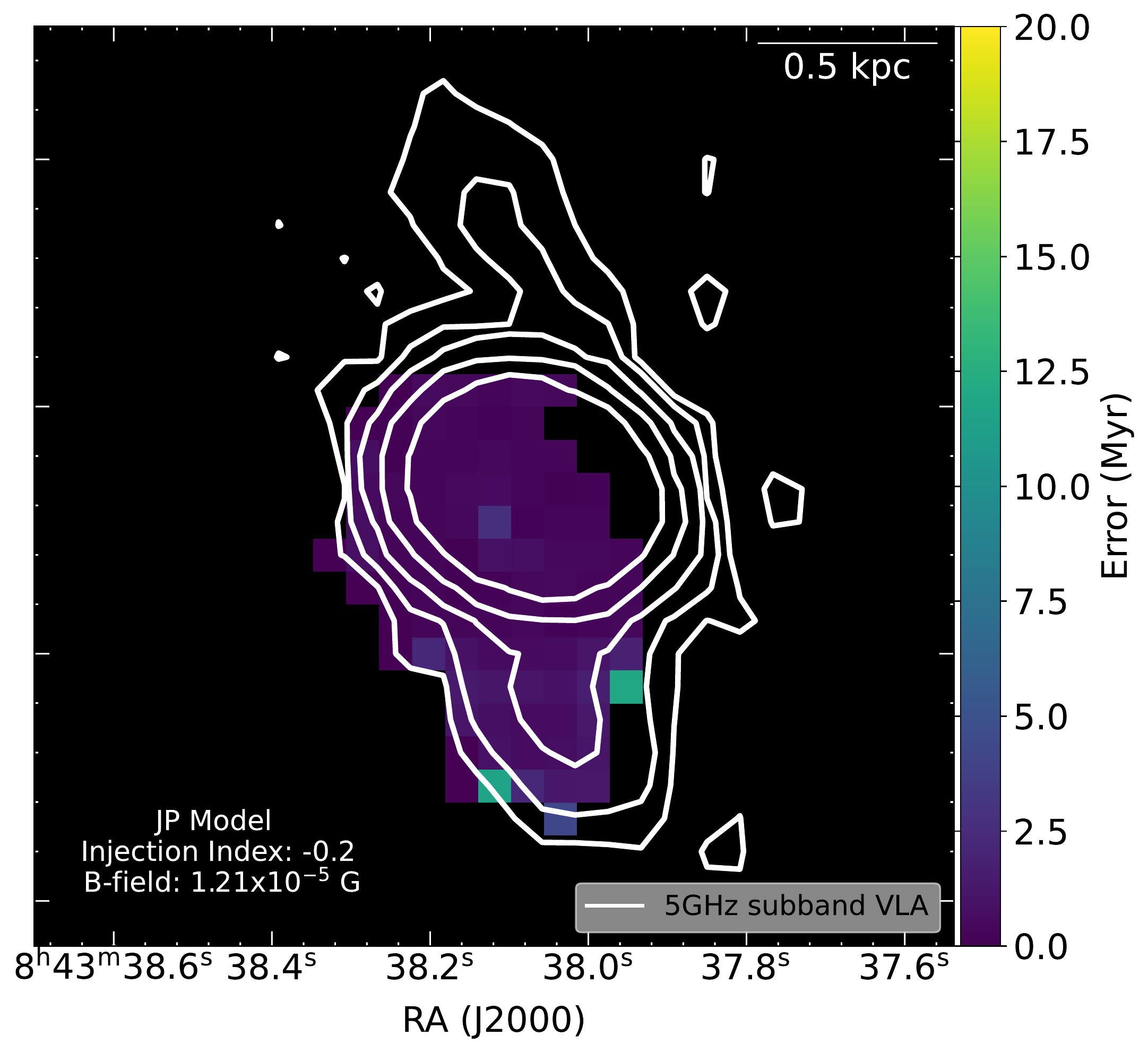}}
\caption{(Left) Spectral age map of the VLA core and the $\sim0.8$ kpc south lobe obtained using the smoothed and regridded images at 1.4, 5.0, and 6.0 GHz. (Centre) The $\chi^2$ map of the fit to the JP model. (Right) The map of the error in estimated spectral age. The mean spectral age of the southern lobe is $12^{+1.7}_{-1.4}$ Myr and the mean spectral age of the core is $2.8^{+0.7}_{-0.5}$ Myr (Note: The yellow regions with high $\chi^2$ in the $\chi^2$ map were excluded from calculating mean). Contour levels are of the 5~GHz VLA subband image (see section \ref{sec:vla-spec-age}) at: $(1,3,9,27,81)\times0.036$~mJy~beam$^{-1}$.}
\label{fig:vla-age}
\end{figure*}

\begin{table*}
\caption{Equipartition estimates. Lobes were assumed to be cuboidal (see Section~\ref{sec:brats}) with a volume filling factor, $\phi=1$, and the ratio of the energy density of ions to electrons, $\eta =1$. The spectral index of each lobe was estimated by taking a spatial average over the lobe from the spectral index maps obtained through BRATS. The standard deviation in this is reported as the error. The sets of images used are the same as in Sections \ref{sec:gmrt-spec-age} and \ref{sec:vla-spec-age}. The constant $c_{12}$ which is a function of spectral index and radio-band frequency cutoffs has been obtained from \citet{Pacholczyk1970}.} 

\label{tab:Table3}
\begin{tabular}{ccccccccc}
\hline
{Lobe } & {Frequency} & {Length} & {Width} & {Total flux density} & {Spectral index $\alpha$} & {$c_{12}$} & {$L_\mathrm{rad}$} & {$B_\mathrm{min}$}\\
{} & {} & {kpc} & {kpc} & {mJy} & {} &{} & {erg~s$^{-1}$} & {G}\\ \hline
GMRT (both lobes) & 735 MHz & 9.0 & 3.1 & $223\pm 22$ & $-0.2\pm 0.2$ & $8.3 \times 10^6$ & $2.8\times 10^{40}$ & $7.6\times 10^{-6}$\\
VLA-south & 5.5 GHz & 0.83 & 0.5 & $1.42 \pm 0.14$ & $-0.53\pm 0.16$ & $1.6 \times 10^7$ & $1.8\times 10^{38}$ & $1.21\times 10^{-5}$\\ \hline
\end{tabular}

The source flux density/ source intensity errors were estimated using the same formula as in Section~\ref{sec:radio}, with the appropriate regions selected.
\end{table*}

\section{Results}\label{sec:results}
Figure~\ref{fig:radopt} shows the total radio intensity image contours at 325~MHZ, 735~MHz, and 5.5~GHZ superimposed on the SDSS \textit{r}-band color map of NGC\,2639. Figure \ref{fig:insets} shows the total intensity image from GMRT, as well as the total intensity images from VLA and VLBA. The new northeast-southwest radio lobes, which were previously not detected in GHz frequency observations with the VLA, are clearly seen in Figures \ref{fig:radopt} and \ref{fig:insets}. The linear extents of each of these lobes are $\sim4.5$ kpc. The PA of the host galaxy disc is $136\degr$ whereas the PA of the northeast-southwest lobes is $34\degr$. The surface discontinuity of the outermost contour of this lobe is $\sim7$ times the \textit{rms} noise. The PA of the VLA north-south lobes is $6\degr$ and that of the east-west lobes is $106\degr$. The surface discontinuities of the outermost contours for the two sets of lobes are $\gtrsim3$ times the \textit{rms} noise. The GMRT image at 325~MHz do not resolve these lobes clearly but detects additional radio emission from the host galaxy itself (see Figure \ref{fig:radopt}). 

\subsection{Episodic Activity and AGN Jet Duty Cycle}
Using multi-frequency arcsec-scale radio data, we have carried out a spectral ageing analysis using BRATS, as described in Section~\ref{sec:brats}. Figures~\ref{fig:gmrt-age} and \ref{fig:vla-age} show the spectral age maps of the GMRT and VLA lobes, respectively. The mean spectral age of the northeast-southwest GMRT lobes is $34^{+4}_{-6}$ Myr, that of the southern VLA lobe is $11.8^{+1.7}_{-1.4}$ Myr, and that of the core is $2.8^{+0.7}_{-0.5}$ Myr. The ages of the southern lobe and core obtained are comparable to the electron lifetime estimates obtained by \citet{Sebastian2019} of $12-16$ Myr and 0.8 Myr respectively. The PA offset of $\sim30\degr$ between the northeast-southwest GMRT lobes and the north-south VLA lobes put together with the spectral ages, which are deduced from the steepness of the radio spectrum, indicate that the two are distinct jet episodes. Thus, the GMRT, VLA, and VLBA data show that NGC\,2639 is a candidate for an AGN with 4 jet episodes. We note that it is possible that the VLBA jet is feeding the east-west VLA lobes \citep[e.g.,][]{Kharb2010}, which could reduce the number of episodes to three. However, in view of the $\sim30\degr$ PA offset between the VLBA jet and the VLA east-west lobes, we will continue to refer to four jet episodes in NGC\,2639. The spectral age results indicate that the jets were launched 9-22 Myr apart with the 9 kpc, northeast-southwest, GMRT jets being launched first followed by the $1.5$~kpc, north-south VLA jets, and the 360 parsec east-west VLA jets, respectively, where the age of the core in Figure~\ref{fig:vla-age} is taken to be indicative of the age of the 360 parsec long east-west VLA jet. Defining the AGN jet duty cycle as $\epsilon=t_\mathrm{on}/(t_\mathrm{on}+t_\mathrm{off})$ \citep[see][]{Clarke1991}, we use the ages of the VLA east-west lobes and north-south lobes to obtain $\epsilon = 11.8/(11.8+9.0) = 0.57$. Similarly, using the ages of the VLA north-south lobes and the GMRT lobes, we obtain $\epsilon = 34.0/(34.0+22.2) = 0.60$. The AGN jet duty cycle in NGC\,2639 is, therefore, $\sim60$\%. These data, however, cannot directly answer whether other AGN activity episodes that might not have produced radio jets occurred in this source. For instance, these data cannot rule out episodic ``wind'' activity that may be unrelated to the radio jets \citep[e.g.,][]{King2015}, leaving an overall uncertainty in the true AGN duty cycle in NGC\,2639.

\begin{figure*}
    \centering
    \subfigure[Moment 0]{\includegraphics[height=5cm]{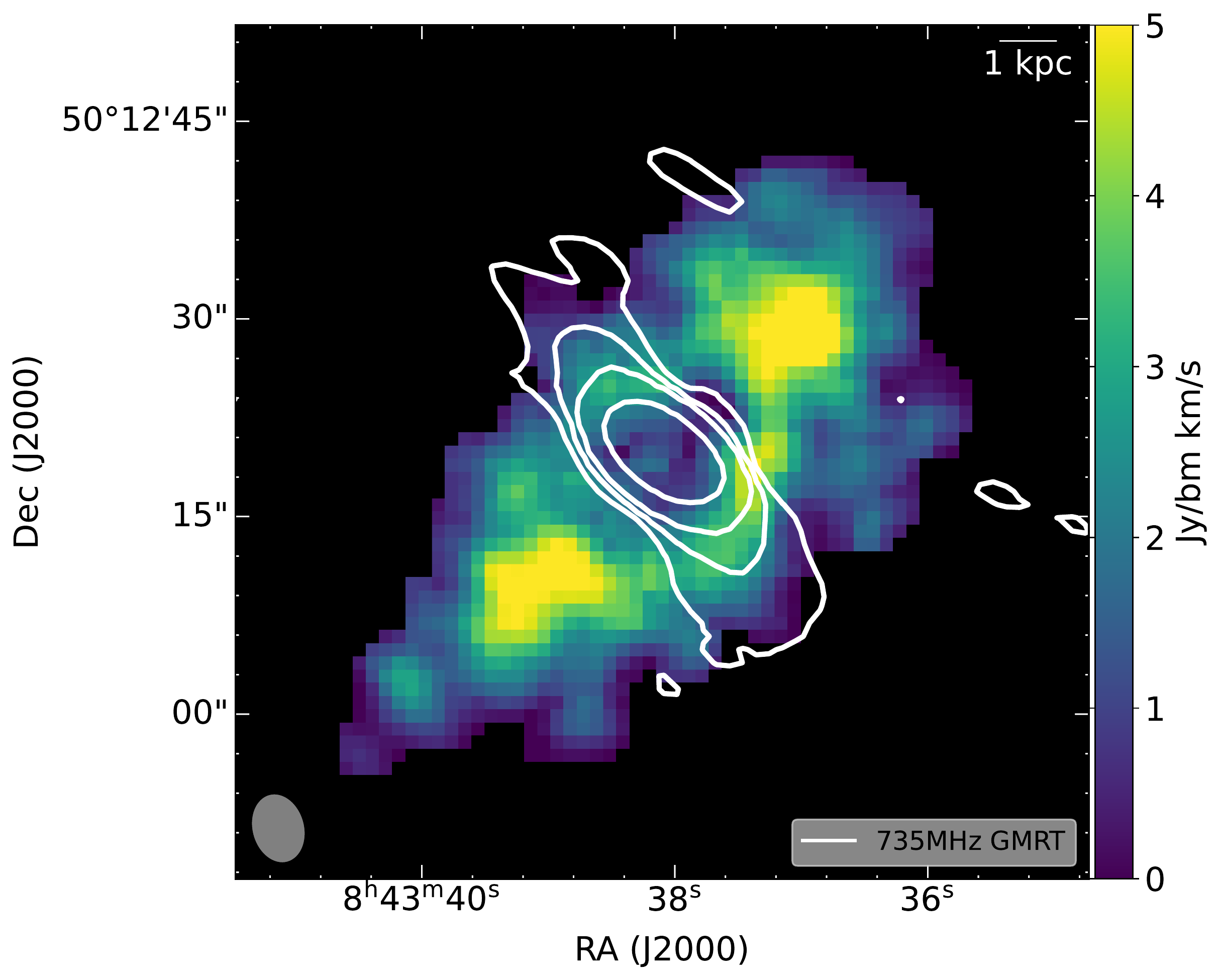}}
    \hfill
    \subfigure[Moment 1]{\includegraphics[height=5cm]{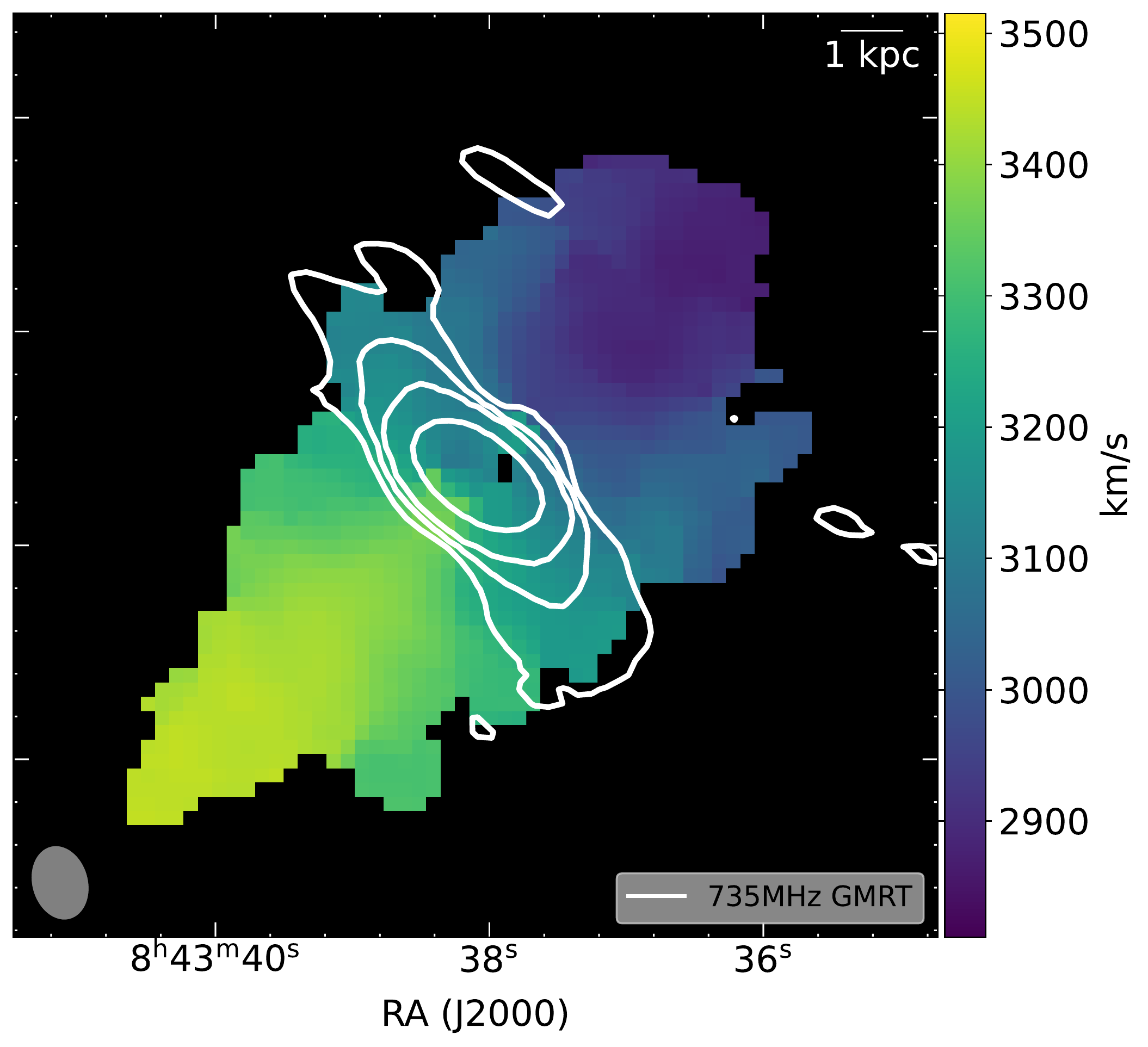}}
    \hfill
    \subfigure[Moment 2]{\includegraphics[height=5cm]{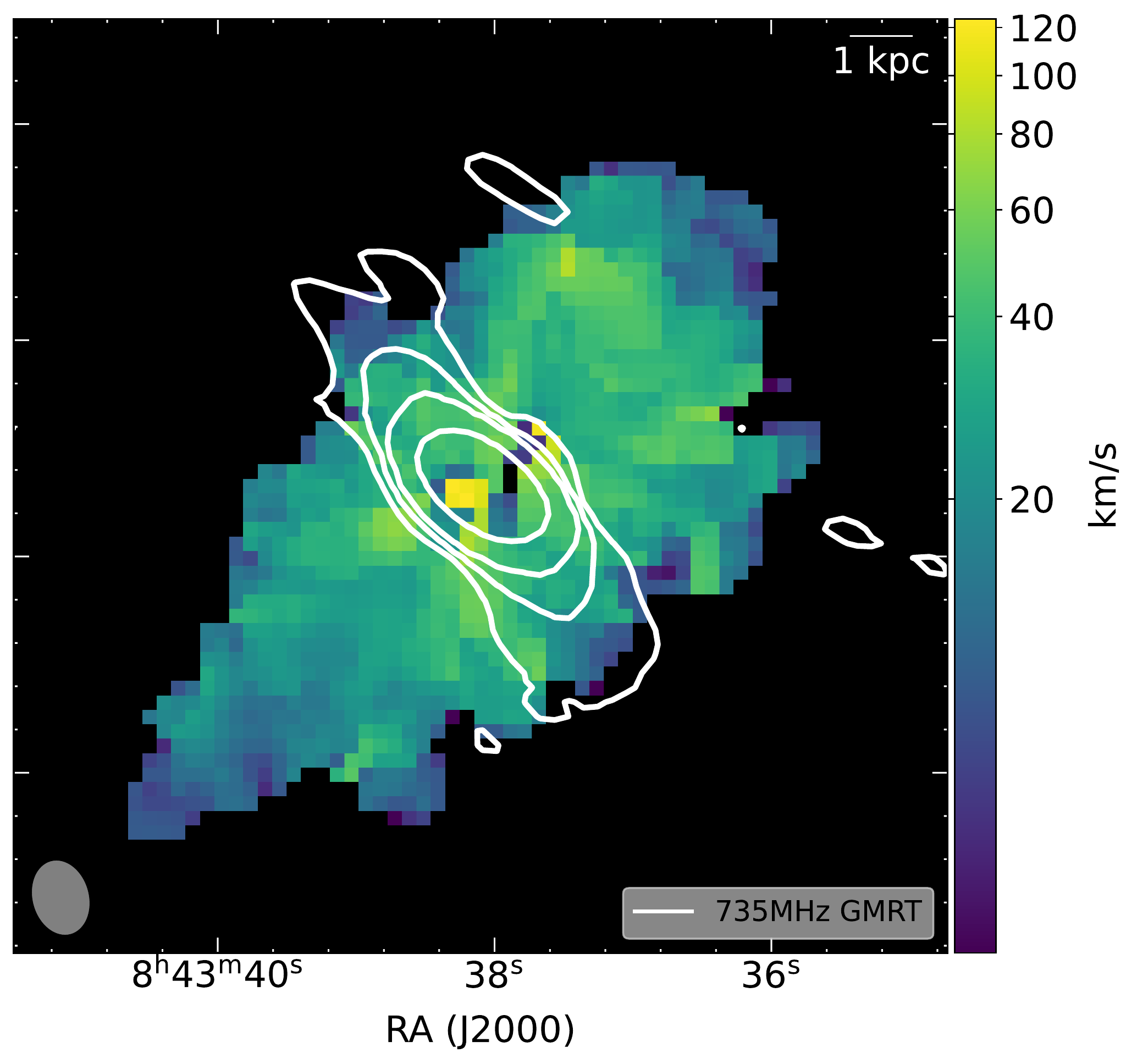}}

\caption{(Left) Moment-0 image of the CO(1-0) molecular gas emission line from the CARMA EDGE survey, overlaid with the 735~MHz GMRT radio contours. This image represents the integrated CO intensity and shows a deficiency of CO(1-0) molecular gas in the central $\sim$6~kpc. (Center) The Moment-1 image representing the velocity of the CO molecular gas. The distribution shows uniform rotation. (Right) The Moment-2 image with the same contour levels, represents the velocity dispersion of the CO molecular gas. Higher velocity dispersion values are observed around the jet edges. Common contour levels are $(1,2,4,32)\times0.6$~mJy~beam$^{-1}$.}
\label{fig:CO}
\end{figure*}

\begin{figure*}
    \centering
    \subfigure[NUV]{\includegraphics[height=7cm]{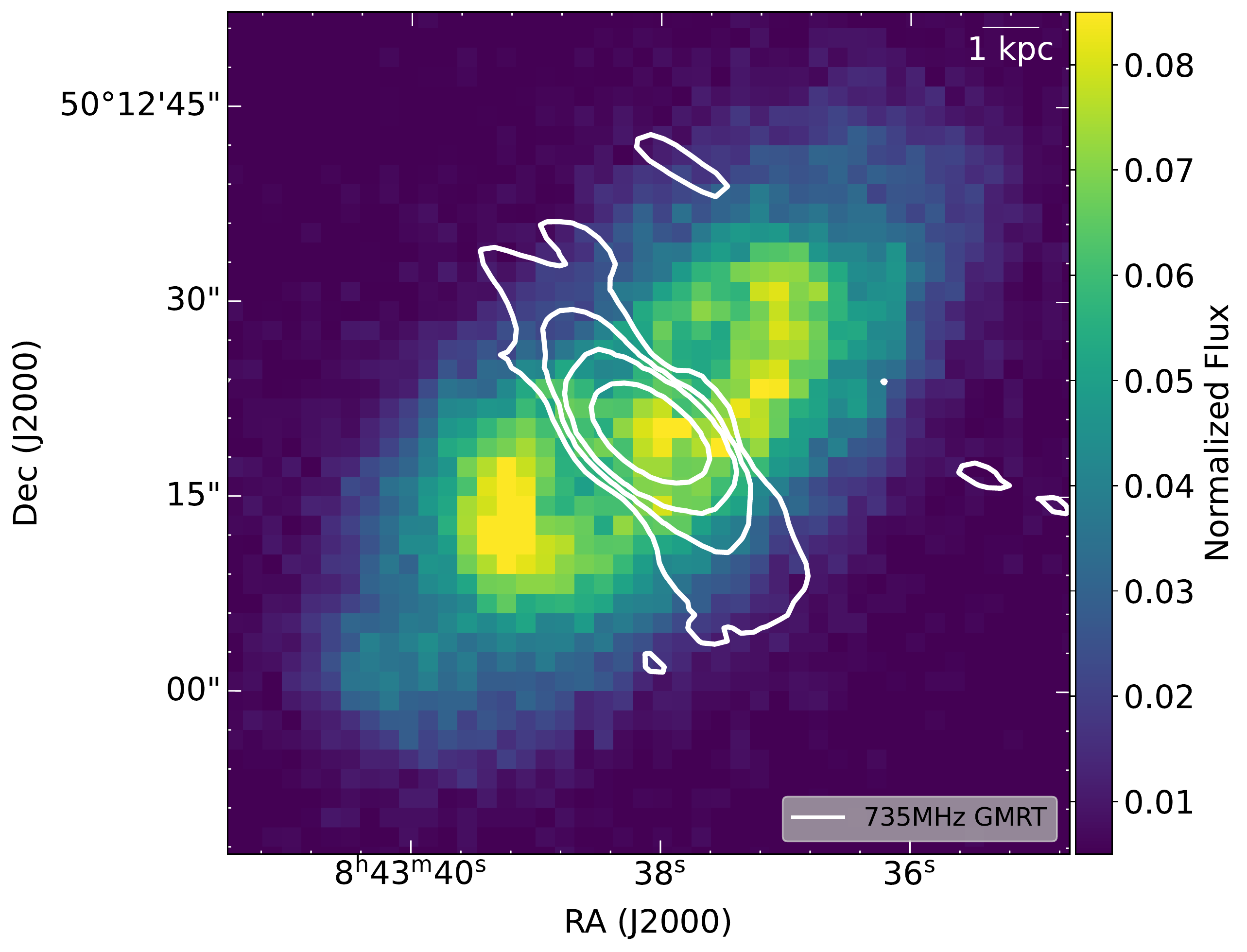}}
    \hfill
    \subfigure[FUV]{\includegraphics[height=7cm]{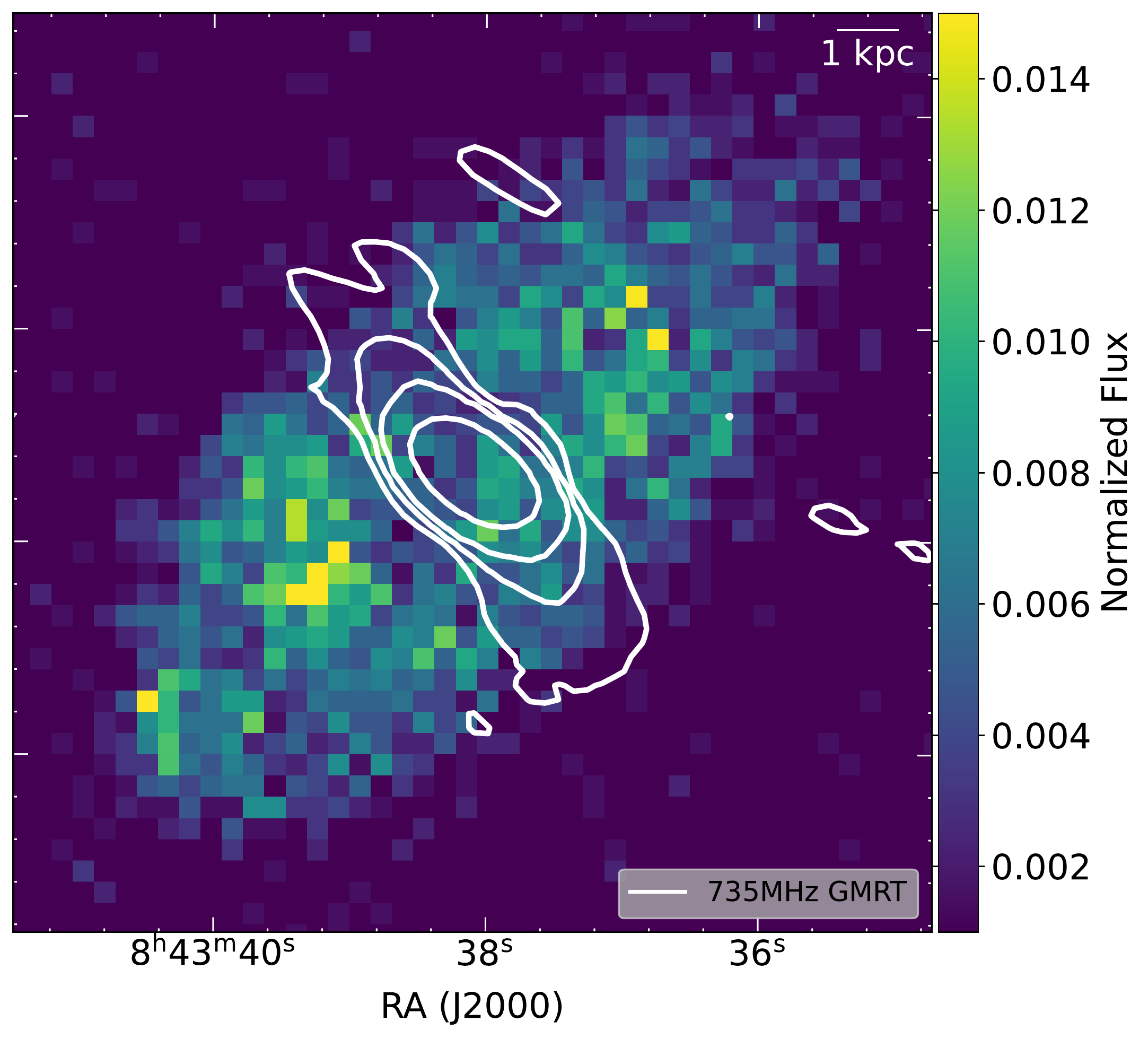}}
    
\caption{(Left) Near-UV and (Right) Far-UV image of NGC\,2639 from GALEX overlaid with the 735~MHz GMRT radio contours at $(1,2,4,32)\times0.6$~mJy~beam$^{-1}$. The central $\sim$6~kpc shows a relative deficiency of NUV emission, barring emission from the AGN, indicating a lack of star formation. The AGN is likely obscured in the FUV image due to dust.}
\label{fig:sfr}
\end{figure*}

\subsection{Looking for the Jet-Gas Connection}\label{sec:feed}
The left panel of Figure~\ref{fig:CO} shows the Moment-0 image of the CO(1-0) molecular gas emission line in NGC\,2639 from the CARMA-EDGE survey \citep{Bolatto2017}. This image represents the integrated CO intensity and clearly shows the deficit of CO(1-0) molecular gas emission in the central $\sim6$~kpc region; the integrated intensity is 5-7 times higher in the ring than in the central regions. Using the same CO(1-0) map, \citet{Ellison2021} obtained the molecular gas surface densities ($\Sigma_{\text{H}_2}$) for NGC\,2639 using the CO-to-H$_2$ conversion factor $\alpha_\mathrm{CO}= 4.3$~M$_\odot$pc$^{-2}$ (K km s$^{-1}$)$^{-1}$. Selecting for AGN and star-forming spaxels using optical line widths from the Calar Alto Legacy Integral Field Area \citep[CALIFA][]{Sanchez2012} survey, they concluded that the molecular gas fractions of central AGN regions are $\sim2$ lower than in star-forming regions, suggesting that the AGN has affected the central molecular gas reservoir. It is worth noting that in comparison to the other three galaxies in their sample, NGC\,2639 exhibited the highest AGN luminosity and a more pronounced lack of molecular gas. While it is possible that the hard ionizing radiation field from the AGN or nuclear star formation can affect the CO spectral line energy distributions, and thereby the low CO transitions, it cannot explain the observed velocity dispersion values (see ahead). Thus, as in \citet{Ellison2021}, we continue to treat the lack of CO(1-0) molecular gas emission as a lack of molecular gas. Here, and in other sections, by molecular gas we primarily mean H$_2$ gas traced by the CO(1-0) emission line. 

The CO(1-0) Moment-1 map of NGC\,2639 (central panel of Figure~\ref{fig:CO}) suggests a rotating molecular gas that appears to be largely regular; clear signatures of turbulence are not directly observed. The Moment-2 map, however, reveals slightly higher velocity-dispersion values (of the order of 100 km~s$^{-1}$) around the GMRT jet edges (see right panel of Figure~\ref{fig:CO}). This could suggest that the jet does impact the molecular gas by injecting turbulent kinetic energy into the medium. If the molecular gas ring is a result of gas being pushed aside due to the radio jet in NGC\,2639, we can estimate the $PV$ (pressure times volume) amount of work required to create such a cavity. Using the typical temperature ($T$) and density ($n$) of the molecular medium of the galactic interstellar medium (ISM), i.e., $T\simeq20$~K, and $n>10^3$~cm$^{-3}$ \citep[see][]{Brinks1990}, and assuming it to be similar in NGC\,2639, we can estimate the molecular gas pressure as $P=nk_BT$, where $k_B$ is the Boltzmann constant. For the molecular gas ring radius of 3~kpc, the volume of the disk-like cavity is $1.25\times10^{66}$~cm$^3$ and the $PV$ work required is $>3.44\times10^{54}$~erg. It is worth noting that the temperature in the ISM may be higher in the vicinity of the AGN, which will increase the required amount of work. However, a much higher temperature will lead to photodissociation of the molecules themselves \citep{Omont2007}.

We can also estimate the jet mechanical powers using the relations derived for low luminosity AGN by \citet{Merloni2007}. Using only the lobe flux densities{\footnote{Lobe flux density is the total flux density minus the core flux density}} at 5~GHz ($\sim$13~mJy), we find that the jet power for the east-west VLA lobes is $7.7\times10^{42}$~erg~s$^{-1}$ and the time-averaged power (for a spectral age of {2.8 Myr) is $6.8\times10^{56}$~erg}. Similarly, for the north-south VLA lobes having a flux density of $\sim$1.4~mJy, the 5.5~GHz jet power is $1.3\times10^{42}$~erg~s$^{-1}$, and its time-averaged power (over 12~Myr) is $4.9\times10^{56}$~erg. Finally, using an average lobe spectral index of $-0.3$ and a lobe flux density of $\sim$16~mJy, the jet power for the north-east - south-west GMRT lobes is $5.7\times10^{42}$~erg~s$^{-1}$, and its time-averaged power {(over 34~Myr) is $6.1\times10^{57}$~erg}. Therefore, only $\sim0.5$\% of the east-west jet power is sufficient to push aside the molecular gas in NGC\,2639; these numbers are $\sim0.7$\% for the north-south jets and $\sim0.06$\% for the north-east - south-west jets, respectively. 

The accumulation of clouds of dense molecular gas around the nucleus of this water maser galaxy seems necessary for the water megamaser to be produced \citep{Martin1989,Krause1990}. However, \citet{Raluy1998} found no apparent relation between the maser luminosity and the surface density or the total content of molecular gas in the innermost kpc-scale galactic regions. The sole correlation found involves the surface density of molecular gas in the inner areas of galaxies, indicating an anticorrelation with the rate of change in maser intensity. This finding suggests that the nuclear jets interacting with matter clouds surrounding the active nucleus at different scales may generate or boost maser emission. \citet{Braatz2003, Pesce2018} have noted that NGC2639 shows evidence for accelerating systemic features, but no high-velocity features have ever been observed for this system.

\subsection{Indications of Star-formation Quenching}
The Galaxy Evolution Explorer (GALEX) data on NGC\,2639 shows a relative deficiency of near-UV (NUV, $\lambda_\mathrm{eff}=2267~{\angstrom}$) emission from the central $\sim6$~kpc of the galaxy (see Figure~\ref{fig:sfr}, left panel), barring the emission from the AGN itself. The far-UV (FUV, $\lambda_\mathrm{eff}=1516~{\angstrom}$) image (Figure~\ref{fig:sfr}, right panel) also shows a similar deficiency of emission but is not as clear as in the NUV, which could be due to the presence of dust. The absence of the AGN in the FUV image, however, would be consistent with its type 2 classification and its obscuration from a dusty torus \citep[see for example,][]{Dewangan2021}. We note that the NUV band is one of the most direct tracers of stars formed over the last $10-200$~Myr \citep{Kennicutt2012}. The central void in the GALEX images, therefore, is consistent with the suggestion of star-formation quenching in the central few kpc in NGC\,2639. We attempt to quantify this further ahead.

The global Schmidt law for star-forming galaxies has been given by \citet{Kennicutt1998} as:
\begin{equation}
\mathrm{\Sigma_{SFR}=(2.5\pm0.7)\times10^{-4}\left(\frac{\Sigma_{gas}}{1~M_\odot~pc^{-2}}\right)^{1.4\pm0.15}~M_\odot~yr^{-1}~kpc^{-2}}
\end{equation}
where the SFR surface density, $\Sigma_\mathrm{SFR}$, can be derived from gas surface density, $\Sigma_\mathrm{gas}$. For NGC\,2639, using $\Sigma_\mathrm{gas}\equiv\Sigma_\mathrm{H_2}$ = 21~M$_\odot$~pc$^{-2}$ for a region of 4.8 kpc, which is the linear size of the antenna beam of the Institut de radioastronomie millimétrique (IRAM) 30~m telescope at the distance of the galaxy \citep{Raluy1998}, $\Sigma_\mathrm{SFR}$ should be  0.0177~M$_\odot$~yr$^{-1}$~kpc$^{-2}$. The SFR surface density can also be computed from the SFR estimates using the following expression:
\begin{equation}
\mathrm{\Sigma_{SFR}=\frac{SFR}{\pi a^2\left(\frac{d}{206265}\right)^2}}
\end{equation}
where the parameter $a$ corresponds to the semi-major axis of the telescope aperture in arcsec and $d$ is the distance to the galaxy in Mpc \citep{Catalan-Torrecilla2015}. Several estimates for SFR have been derived for NGC\,2639 in the literature. Table \ref{tab:Table4} provides the estimates of $\Sigma_\mathrm{SFR}$ obtained using different SFR tracers. The telescope details for individual tracers are also provided. These estimates of $\Sigma_\mathrm{SFR}$ lie below the global Schmidt best-fit line for star-forming galaxies by a factor of $5-18$, consistent with star formation quenching in NGC\,2639. \citet{Kennicutt1998} have noted that the scatter in this relation is $\pm0.3$ dex with individual sources deviating by as much a factor as 7. These results are consistent with the GALEX NUV image of NGC\,2639, showing a deficit in recent star-formation in the central $\sim6$~kpc region.  

\begin{table*}
\caption{Estimates of star formation rate in NGC~2639 using different tracers}
\label{tab:Table4}
\begin{tabular}{cccccc}
\hline
SFR & Telescope & SFR indicator & Aperture & $\Sigma_\mathrm{SFR}$ &
Reference \\
(M$_\odot$~yr$^{-1}$) & & & ($\arcsec$) & (M$_\odot$~yr$^{-1}$~kpc$^{-2}$) & \\
\hline
0.92 & Nickel 1.0 m telescope at Lick Observatory & H${\alpha}$ & 73.5 & 0.00099 & (1) \\ 
1 & Spitzer Space Telescope & IR & 40 & 0.0036 & (2) \\
0.57 & Calar Alto 3.5 m telescope & H${\alpha}$ & 36 & 0.0026 & (3) \\
\hline
\end{tabular}

References: 
(1) \citet{Theios2016} using the relation, $\mathrm{SFR~(M_\odot \,yr^{-1})= 5.37 \times 10^{-42}~L_{H{\alpha}}~(erg\,s^{-1}) }$, where L$_\mathrm{H{\alpha}}$ = 10$^{41.23}$~(erg\,s$^{-1}$) for NGC\,2639 \\
(2) \citet{Sebastian2019} using the CLUMPYDREAM code \\
(3) \citet{Catalan-Torrecilla2015} using the H${\alpha}$ line luminosity from the CALIFA survey  
\end{table*}

\section{Discussion}\label{sec:discussion}
NGC\,2639 is a member of the non-interacting spiral (NIS) galaxy sample considered for a stellar population modeling study by \citet{Zaragoza2018} and has been identified as such by \citet{GildePaz2007}. The stars of NGC 2639 are relatively undisturbed and are in uniform motion, as seen in the line of sight velocity map obtained from the CALIFA survey \citep{PyCASSO2017}. According to
\citet{Sebastian2019}, the multiple radio lobes seen in NGC\,2639 are due to minor mergers that did not disrupt the morphology of the host galaxy. The misaligned jets could be the result of new accretion disks formed from mergers, with jet directions conserving the angular momentum of the inflowing gas \citep[e.g.,][]{Kharb2006,Volonteri2007}. This scenario is consistent with NGC\,2639 having a large bulge component surrounding the nucleus \citep{Cox2008} as gravitational forces and torques that result from mergers disrupt the orbital path of stars causing randomized bulge orbits. Thus, if the minor merger scenario were true, the spectral age results of the multiple lobes indicate that minor mergers occurred every $9-22$~Myr apart in the last $\sim30$~Myr. 

The expected minor merger rate for a galaxy like NGC\,2639 (redshift of 0.01113 and stellar mass of $1.48\times10^{11}$~M$_{\sun}$) is $\sim13$~Myr following the work of \citet{Conselice2022}, who used observational data from the REFINE survey. We used their minor mergers best fit line (with stellar mass ratios of 1:10) to obtain this estimate. The estimate of $\sim10$~Myr also matches the estimates obtained via theoretical studies as well as galaxy-merger simulations \citep{Hopkins2010,Capelo2015}. It would therefore be fair to conclude that at least three minor mergers have taken place in the lifetime of NGC\,2639. Each of these mergers may have resulted in the formation of a new accretion disk with no memory of the previous accretion disk direction, primarily driven by the angular momentum of the infalling material itself \citep[e.g.,][]{Kharb2006}. Accretion through these disks would have resulted in the several differently-oriented jet episodes that are observed in NGC\,2639. As noted in Section~\ref{sec:feed}, each of these jet episodes have sufficient mechanical power to displace the CO and H$_2$ molecular gas from the central few kpc of the host galaxy.

The early-type galaxy and LINER, NGC\,1266, shows the presence of a CO molecular outflow, no signatures of galaxy interactions, and a possible radio jet at 1.4~GHz \citep{Alatalo2011}. There is also a centrally concentrated molecular component which is different from the case of NGC\,2639, where instead, a central deficiency is observed. \citet{Alatalo2011} suggest that the jet in NGC\,1266 is sufficient to  drive the molecular outflow using only 2\% of its total power at 1.4~GHz. We note, however, that multi-resolution, multi-frequency radio observations are required to truly rule out the existence of multiple jet episodes in NGC\,1266. \citet{Nesvadba2021} detected a CO(1-2) molecular gas ring through ALMA observations in the nearby spiral galaxy J2345$-$0449 with large kpc-scale radio jets. Interestingly, the inner radius of the CO gas ring corresponds to $4.2\times2.2$~kpc, very similar to what is observed in NGC\,2639. Moreover, they find that the molecular gas outflow in J2345$-$0449 has a kinetic energy of $1.3\times10^{57}$~erg. Again, only a small fraction of the jet kinetic power in J2345$-$0449 (and as it happens in NGC\,2639) can suffice to drive molecular gas outflows. It is worth noting that the radio source in J2345$-$0449 is also a restarted double-double radio galaxy.

The simulations performed by \citet{Mukherjee2016, Mukherjee2017} show that low-power jets ($P_\mathrm{jet}\lesssim 10^{44}$ erg s$^{-1}$) have a more pronounced impact on ISM evolution in comparison to high-power jets ($P_\mathrm{jet}\gtrsim 10^{45}$ erg s$^{-1}$) due to their longer trapping time and therefore the constant stirring of the turbulent ISM, ultimately leading to a suppression of star formation. The low-power jets also evacuate $\gtrsim 1$~kpc cavities, with cavity sizes increasing with decreasing power. The simulations show that only a small percentage of the dense gas mass is ejected by jets. However, the injection of turbulent kinetic energy by the jet into the ISM may temporarily suppress star formation in the remaining gas, as observed by \citet{Alatalo2015}. This suppression is temporary as the ejected mass eventually falls back into the ISM and may become available for star formation after a few tens of Myr. 

Recent simulations of \citet{Mukherjee2018, Meenakshi2022} have demonstrated that jet-ISM coupling is sensitive to the relative orientation of the jet w.r.t the gas disk, as well as the age of the jet. It is stronger for the jets, which are oriented at $\leq45\degr$ w.r.t the gas disk and young ($\leq2$ Myr). Thus, the absence of CO(1-0) emission in the inner 6~kpc region of NGC\,2639, the relatively small values of $\Sigma_\mathrm{SFR}$ observed compared to the Schmidt law for star-forming galaxies, and the high CO(1-0) velocity dispersion seen in the central regions, together paint a complex picture of “negative AGN feedback” in NGC\,2639. Given the directional nature of the jets in NGC\,2639, we speculate that the jets restarting repeatedly on timescales comparable to the fallback time of the gas is an important factor in launching multiple local outflows, and maintaining a turbulent ISM, regulating star formation. NGC\,2639 is thus, another potential candidate to add to the increasing number of observed instances \citep[e.g.][]{Alatalo2015, Venturi2021, Girdhar2022} where low-power radio jets have been identified as a significant source of feedback.

\section{Conclusions}\label{sec:summary}
The Seyfert galaxy NGC\,2639 exhibits four episodes of AGN jet activity as evidenced by 735~MHz, 5.5~GHz, and 8.3~GHz frequency observations via the GMRT, VLA, and the VLBA telescopes, respectively. Using the spectral ageing software BRATS, we derive the ages of the three pairs of lobes to be respectively, $34^{+4}_{-6}$ Myr, $11.8^{+1.7}_{-1.4}$ Myr, and $2.8^{+0.7}_{-0.5}$ Myr, with the GMRT lobes being the oldest (we did not derive an age for the VLBA jet). Using the ``on'' and ``off'' times of these jets/lobes, the AGN jet duty cycle in NGC\,2639 is $\sim60$\%. NGC\,2639 also shows a deficiency of molecular gas in its central $\sim6$~kpc region. Less than 1\% of the jet mechanical power for each of the jet episodes taken individually, is sufficient to move the molecular gas. However, given the directionality of each jet episode, the creation and maintenance of a doughnut-shaped hole in the molecular gas in the galactic center likely required multiple jets restarting on timescales comparable to the fallback times of ejected molecular gas. Like the CO(1-0) emission line image, the GALEX NUV image also shows a relative deficiency of star formation in the last 200 Myr in the inner $\sim6$~kpc region. Additionally, the SFR surface density is lower by a factor of $5-18$ compared to the global Schmidt law of star-forming galaxies. The results suggest that the central regions of NGC\,2639 are influenced by multiple low-power jets, which could be playing a key role in regulating star formation. This would make NGC\,2639 a case of a radio-quiet AGN showing episodic jet activity and possible signatures of jet-driven AGN feedback.



\section*{Acknowledgements}
We thank the anonymous referee for their insightful suggestions that have improved this manuscript.
We thank the staff of the GMRT that made these observations possible. GMRT is run by the National Centre for Radio Astrophysics of the Tata Institute of Fundamental Research. PK, RK, JB, SS and SM acknowledge the support of the Department of Atomic Energy, Government of India, under the project 12-R\&D-TFR-5.02-0700. The National Radio Astronomy Observatory is a facility of the National Science Foundation operated under cooperative agreement by Associated Universities, Inc.

\section*{Data Availability}
The data underlying this article will be shared on reasonable request to the corresponding author.


\bibliographystyle{mnras}
\bibliography{ms} 



\bsp	
\label{lastpage}
\end{document}